# Understanding high ordering temperature in $Gd_6FeBi_2$ magnet: critical behavior, electronic structure and crystal-field analysis

Guoming Cui[a], Huan Ma[b], Kam Wa Wong[c], Chan Hung Shek[b], Guangcun Shan[d],*, Jiliang Zhang[e],*

[a]Department of Materials Science and Engineering, Henan Institute of Technology, Xinxiang 453003, China

[b]Department of Materials Science and Engineering, City University of Hong Kong, Kowloon, Hong Kong

[c]Department of Physics, City University of Hong Kong, Kowloon, Hong Kong

[d]School of Instrument Science and Opto-electronics Engineering & Institute of Quantum Sensing, Beihang University, Beijing 100083, China

[e]Department of Energy & Materials Engineering, Dongguk University-Seoul, Seoul, 04620 South Korea

*Corresponding author: gcshan@buaa.edu.cn; jiliangz@dongguk.edu

**Abstract**: $Gd_6FeBi_2$ is reported as the only one room-temperature magnet with a Curie temperature ($T_c$) of ca. 350 K among more than hundreds of compounds with its structural type, which makes it more attractive in potential applications. To reveal the origin of such high ordering temperature, critical behaviors, electronic structure and crystal-field effects of $Gd_6FeBi_2$ are investigated in this work. The short-range Gd-Fe ferrimagnetic interaction is supported by the non-Curie-Weiss paramagnetic behavior, crystal and electronic structure analyses, in agreement with previous DFT calculations. Unlike the strong *TM-TM* exchange interactions, the Gd-Fe exchange interaction shows limited influence on the critical exponents determined by long-range exchange interactions, which seems a common feature in *RE-TM* based alloys without *TM-TM* exchange interactions. However, the strong Gd-Fe hybridization reduces the influence of vibronic couplings on the short-range exchange interaction and thus allows a high $T_c$. The broadening or splitting mechanism of Gd 4*f*-electron bands is addressed based on crystal-field analysis and likely another factor for elevated $T_c$ in $Gd_6FeBi_2$ and Gd-based compounds with non-magnetic elements. Different magnetic behaviors among isostructural compounds, and the relationship between the band splitting and crystal-field effects is also discussed.

## 1. Introduction

Most heavy rare-earth (*RE*) metals possess very large magnetic moments ideal for many magnetic applications, but the low magnetic ordering temperature inhibits their application in practice. To overcome this drawback, many efforts are made to develop *RE*-based *RE-TM* (*TM*, transition metal) intermetallic compounds, because the strong *RE-TM* or *TM-TM* magnetic interactions can effectively elevate the ordering temperature compared with the indirect 4f-4f interactions between *RE* atoms, while the small magnetic moment of *TM* will not change the net magnetic moment of these compounds significantly. Another chemical element, generally from the p-block of main groups, is usually added to these *RE-TM* systems to produce even greater diversity of new compounds and structures, and to improve properties as well.

A new family of Fe$_2$P type with the general formula $RE_6TMX_2$ ($RE$ = Gd-Tm, and Lu; $TM$ = Mn, Fe, Co, Ni and Ru; and $X$ = Al, Ga, Sn, As, Sb, Bi, S, Se and Te) firstly discovered in 2003 by different research groups independently is a typical example of such strategy [1-5]. These compounds show fruitful magnetic behaviours as characterized by multiple magnetic transitions and the absence of magnetic moment from $TM$ in many members [6-8]. Some members also show large magnetocaloric effects around their Curie temperatures ($T_c$) [9-12]. However, the reported magnetic ordering temperature of all members are lower than room temperature except Gd$_6$FeBi$_2$ [11], which has an ordering temperature up to 350 K, higher than that of most Gd-based intermetallic compounds [13].

Due to the great potential for practical application, many efforts are made to understand the origin of high ordering temperature or strong magnetic interactions in these Gd-based compounds with high $T_c$, e.g. Gd$_5$Si$_4$ (336 K) and Gd$_4$Bi$_3$ (340 K) [14, 15]. A typical explanation for the compounds with non-magnetic elements (NMEs) is the Ruderman-Kittle-Kasuya-Yosida (RKKY) interaction, where the f-shell electron spins couple with those of a neighbouring Gd atom through the conduction electrons. Such indirect exchange interaction can induce the fluctuation of $T_c$ depending on the conduction electrons and distance between neighbouring magnetic atoms, but it is difficult to elevate the Curie temperature significantly because of the nature of indirect interactions. Later it is realized that the high $T_c$ can be induced by extra magnetic moments from the localized/polarized 5$d$ electrons in Gd [16], or extra magnetic moments from NMEs (e.g. Bi in Gd$_4$Bi$_3$) due to strong interactions or polarizations [17]. Very recently, the low-temperature structure of Gd$_6$FeBi$_2$ was determined from single crystals, and the DFT calculations based on the structural model did show the magnetic moments from 5*d* electrons of Gd and a ferrimagnetic interaction between Gd and Fe, accounting for the high ordering temperature in $Gd_6FeBi_2$ [18]. However, multiple magnetic transitions (and very likely structural transitions) exist in this family and some may induce a significant change of interactions between atoms. Thus, to understand the origin for the high $T_c$, it is more proper to use the data from the high-temperature magnetic phase for analysis.

In this work, we attempt to reveal the magnetic exchange interactions from the critical behaviours of the high-temperature magnetic phase and understand the chemical and magnetic interactions in the compound from its electronic structure and crystal-field analysis, so that the structural landscape for the enhanced magnetic exchange interactions can be constructed.

## 2. Experimental

$Gd_6FeBi_2$ ingots were fabricated by arc melting the mixtures of pure constituent elements Gd (99.9 Wt. %), Fe (99.95 Wt. %), and Bi (99.9 Wt. %) under a Ti-getterred argon atmosphere. A small extra amount of Bi (0.4%, 0.6%, 0.8%, 1.0% and 1.2%, respectively) were added into samples to compensate for the loss of Bi due to vaporization during arc melting. All ingots were remelted for four times to improve compositional homogeneity. After melting, all ingots were sealed into high vacuum quartz tube and then annealed at 1073 K for 10 days. Finally, the sealed quartz tube was quenched into ice water. Phase identification was conducted on a Philips X'pert X-rays diffractometer (Cu Kα, λ=0.15406 nm). Isothermal magnetisation were measured using a LakeShore VSM with a maximum field of up to 50 kOe. The applied magnetic fields ($H_E$) were corrected for the demagnetization to get the internal field $H=H_E-NM(T,H_E)$, where N is the demagnetion factor and M is the measured magnetization. Demagnetization factors were calculated from low-field magnetization following the method in given in reference [19]. To further reduce the effects of demagnetization, only data obtained from the field above 0.5 T were used for analysis. Experimental electronic structures were measured by X-ray photoelectron spectroscopy (XPS) using an ULVAC-PHI 5802 spectrometer equipped with a monochromatized Al Kα radiation (hυ=1486.6 eV) in vacuum of about $10^{-9}$ mbar. All samples were sputtered by Ar ion to remove the $O_2$-contiminated surface before the measurement.

3. **Scaling analysis**

A variety of physical systems exhibited critical phenomena, in which the physical behaviors can be well described using a set of parameters. And the second order magnetic transition is a well-known critical point. For such magnetic systems, the low-temperature magnetization and high-temperature susceptibility can be well described using a scaling rule constructed by critical exponents. According to the scaling hypothesis, the mathematical definitions of these critical exponents for magnetic systems can be described as follows [20]:

$$M_S(T) = M_0(-\varepsilon)^\beta, \varepsilon < 0 \quad (1)$$

$$\chi_0^{-1}(T) = \Gamma\varepsilon^\gamma, \varepsilon > 0 \quad (2)$$

$$M = DH^{1/\delta}, \varepsilon = 0 \quad (3)$$

where:

$\varepsilon = (T - T_C)/T_C$, the reduced temperature; $M_0$, $\Gamma$, $D$, the critical amplitudes;

$\beta$, a critical exponent associated with the spontaneous magnetization $M_S$;

$\gamma$, a critical exponent associated with the initial magnetic susceptibility $\chi_0$;

$\delta$, a critical exponent associated with the critical isothermal magnetization at $T_c$.

These exponents are not independent of each other, but correlated by some relationships. And finally a scaling hypothesis, describing the magnetic equation of state, is achieved. Using the scaling hypothesis the magnetization can be expressed as:

$$M(H, \varepsilon) = |\varepsilon|^\beta f_\pm \left(\frac{H}{|\varepsilon|^{\beta+\gamma}}\right) \quad (4)$$

where $f_\pm$ are regular analytical functions with $f_+$ for $\varepsilon > 0$, and $f_-$ for $\varepsilon < 0$. In terms of scaled magnetization $m \equiv |\varepsilon|^{-\beta} M(H, \varepsilon)$ and scaled field $h \equiv |\varepsilon|^{-(\beta+\gamma)} H$, the Eq. (4) can be written into the more familiar form:

$$m = f_\pm(h) \quad (5)$$

It is clear that the equation above implies that the scaled $m$ plotted as a function of the scaled $h$ will fall onto two different universal curves described by $f_+$ and $f_-$ respectively, for true scaling relations and right choice of these critical exponents.

Although the exponents generally show universal properties in the asymptotic region ($\varepsilon \to 0$), various systemic trends or crossover phenomenon are often observed, mostly due to the presence of various competing couplings and/or disorder. In the present case, $Gd_6FeBi_2$ shows non-Curie-Weise paramagnetic behaviors, in contrast to the HCP Gd with clearly Cuie-Weise paramagnetic behaviors above magnetic transition [21]. Therefore it is useful to introduce the temperature-dependent effective exponents for $\varepsilon \neq 0$, which are defined as [20]:

$$\beta_{eff}(\varepsilon) = \frac{d[ln M_S(\varepsilon)]}{d(ln\varepsilon)} \quad (6),$$

$$\gamma_{eff}(\varepsilon) = \frac{d[ln \chi_0^{-1}(\varepsilon)]}{d(ln\varepsilon)} \quad (7)$$

These effective exponents are general non-universal, but approach universal exponents in the asymptotic limit.

**4. Results and discussion**

Among the synthesized samples, the one with an extra 0.8% Bi has the best phase purity. The XRD pattern of $Gd_6FeBi_2$ at room temperature is similar to the simulated XRD pattern generated from structure parameters obtained by single crystals despite the slight shift due to different temperatures (see Fig 1) [18]. All reflections in XRD pattern of $Gd_6FeBi_2$ can be well indexed according to the $Fe_2P$-type structure, and gives the unit cell of a = 8.366(5) Å and c = 4.243(4) Å, in good agreement with previous reports [5, 11]. RKKY indirect exchange interaction, which is often common in *RE*-based compounds without other NMEs, can induce the fluctuation of $T_c$ depending on the concentration of conduction electrons and distance between neighbouring magnetic atoms. Our recent work shows that exchange interaction in $Gd_6CoTe_2$ is dominated by the RKKY mechanism, while extra short-range exchange interactions exist in $Gd_6FeBi_2$ accounting for the high $T_c$ [18]. To understand these exchange interactions, critical behaviors around the $T_c$ is to be first investigated.

**4.1 Critical behaviors**

Near the transition, the Landau theory of second-order phase transition suggests that the free energy G of a magnetic system can be expanded in the powers of its order parameter *M* in the following form [22]:

$$G(T,M) = G_0 + \frac{1}{2}A(T)M^2 + \frac{1}{4}B(T)M^4 + \cdots \cdots - MH \quad (8)$$

where coefficients A and B are temperature-dependent parameters, and the last term describes the energy of spins in an external field *H*. Generally the higher order items can be neglected in practice due to their small values. In the case of equilibrium, the energy is minimized ($\frac{\partial G}{\partial M} = 0$), which leads to the magnetic equation of states in the form of Arrott formula [23]:

$$\frac{M}{H} = A + BM^2 \quad (9)$$

Therefore the $M^2$ versus H/M curves should be straight lines, of which the intercept on the H/M axis determines the ordering transition. As shown in Fig 2a, the Arrott plots of $Gd_6FeBi_2$ exhibit positive slope, which indicates that the magnetic transition is second-order according to the criterion suggested by Banerjee [24]. It is also noted in Fig 2a that the plots of $Gd_6FeBi_2$ show slightly downward from the linearity, because the Arrott plot is actually based on the mean-field approach without considering microscopic exchange interactions and fluctuations in magnetic systems.

Compared with the mean-field theory, both Ising and Heisenberg models include the exchange interaction between spins (short-range/direct interactions). And both models can also give critical exponents for the well-known modified Arrott (MA) relation which can be expressed as [25]:

$$(\frac{M}{H})^{1/\gamma} = A\varepsilon + BM^{1/\beta} \quad (10).$$

Based on critical exponents $\beta$ and $\gamma$ predicted by 3D-Ising and 3D-Heisenberg models (see Table 1) [26], MA plots of $Gd_6FeBi_2$ are drawn in Fig 2b and 2c, respectively, and all plots seem curved upward deviating obviously from the linearity.

According to Eq. (10), the linear extrapolation of the high field portions of the isotherms will give an intercept on both $M^{1/\beta}$ and $(\frac{M}{H})^{1/\gamma}$ axis, from which the spontaneous magnetization $M_S$ and the inverse initial susceptibility $\chi_0^{-1}$ can be calculated. Using these calculated data, new critical exponents can also be obtained based on Eq. (1) and (2). The best values of $\beta$ and $\gamma$ obtained by fitting Eq. (10) should be calculated self-consistently with the values yielded by fitting Eq. (1) and (2) to the extrapolated data. To get the best values, we performed according to the idea which was introduced in literature [27], and then got the best values fitting our data. Previous work shows that effective exponents converge approaching to the universal exponents only when $\varepsilon < 0.1$ [26]. Thus our estimations on critical exponents were performed using data in the range. After several cycles, the exponents converged into stable values. The values obtained after 30 cycles are $\beta$ = 0.441(8), $\gamma$ = 1.098(12) for $Gd_6FeBi_2$. Evidently, the MA plot of $Gd_6FeBi_2$ exhibit good linearity around $T_c$ as shown in Fig 2d. The spontaneous magnetization $M_S$ and the inverse initial susceptibility $\chi_0^{-1}$ obtained using these exponents are then plotted as a function of temperature in Fig 3a. The fitting of these values to Eq. (1) gives $\beta$= 0.439(6), $T_c$ = 347.9(1) K, and to Eq. (2) gives $\gamma$ = 1.107(10), $T_c$ = 348.1(1) K. These values are also listed in Table 1. It is distinct that these values are closer to those given by mean-field model rather than 3D-Ising or 3D-Heisenberg model.

Using determined $M_S$ and $\chi_0^{-1}$ by above method (see Fig 3a), the critical exponents and $T_c$ of high precision can be obtained by following the Kouvel-Fisher (KF) method [28], which is described by an alternative form of Eqs. (1) and (2):

$$M_S(\frac{dM_S}{dT})^{-1} = (T - T_C)/\beta \quad (11)$$

$$\chi_0^{-1}(\frac{d\chi_0^{-1}}{dT})^{-1} = (T - T_C)/\gamma \quad (12).$$

According to the method, in critical regions, the plots of $M_S(\frac{dM_S}{dT})^{-1}$ vs. $T$ and $\chi_0^{-1}(\frac{d\chi_0^{-1}}{dT})^{-1}$ vs. $T$ yield straight lines with slopes $1/\beta$ and $1/\gamma$ respectively, and intercepts of such fitted straight lines on their $T$ axis equal to $T_c$. The most important advantages of the KF method is: no previous knowledge of $T_c$ is required, and it provides a consistency condition for $T_c$, namely, the fitting of both plots should give the same value of $T_c$. The KF plots of $Gd_6FeBi_2$ is shown in Fig 3b, and estimated critical exponents and $T_C$ by fitting these straight lines are: $\beta$=0.446(6), $T_c$=348.2(1) K and $\gamma$=1.092(10), $T_c$=348.0(1) K.

All these critical exponents estimated from above methods, together with predicted theoretically values from different models, are listed in Table 1. It is evident that values of critical exponents and $T_c$ calculated using both methods match reasonably well, which indicates that these estimated values are self-consistence. However, it is also clear that these values do not match the conventional universality classes. Thus it is necessary to verify whether these critical exponents can produce the scaling equation of state. Taking the values of critical exponents and $T_c$ listed in Table 1, the plots of scaled m against scaled h are drawn in Fig 4a. These plots depict the two different universal curves distinctly as predicted by the scaling equation Eq. (5), which indicates the reliability of the calculated critical exponents and $T_c$. The inset of Fig 4a shows the same plot on log-log scale, which suggests the converging of two curves towards to $T_c$.

A more rigorous way to confirm whether the estimated values of critical exponents and $T_c$ are reliable and whether the isotherms taken in the critical region obey the scaling equation of state, is to analyze the experimental data in terms of the well-established asymptotic form of the scaling equation given by [29]

$$\frac{h}{m} = \pm a_\pm + b_\pm m^2 \quad (13)$$

where plus and minus signs have the same meaning as that in Eq. (4), *a* and *b* are two scaling parameters for Gibbs potential concerning both thermal and magnetic field effects. Then the coefficients in the equation can be related to the critical amplitudes in Eqs. (1) and (2) as

$$(a_-/b_-)^{1/2} = M = m_0 \quad (14)$$

$$a_+ = \Gamma = h_0/m_0 \quad (15)$$

It is significant that these critical amplitudes $m_0$ and $h_0/m_0$ can be obtained by intercepting the universal curves with $m^2$ and $\frac{h}{m}$ axis in the plots of $m^2$ vs. $\frac{h}{m}$ respectively, as shown in Fig 4b. These values are determined as: $m_0$=169(6) emu/g, and the effective exchange interaction field $h_0$=11.6(2) kOe.

### 4.2 Magnetic exchange interaction and electronic structure

In *RE*-based compounds with NMEs, the RRKY interaction is generally the dominant long-range interaction, but other long-range forces like dipolar interaction can also influence the critical fluctuations of magnetization. According to the criterion given by Tadaoff et al [30], such long-range forces can be neglected in case

$$|\varepsilon| \gg [\mu M_S(0)/k_B T_C]^{1/\beta(\delta-1)} \equiv t \quad (16)$$

where $\mu = g\mu_B S$ is the moment of per spin, $k_B$ is Boltzmann constant, and the saturation magnetization $M_S(0)$ can be calculated form equation (1) as 1600 Oe. Then using equation (16) $t$ is estimated as $2.5 \times 10^{-3}$, much smaller than the range of $|\varepsilon|$. Thus the dipolar interactions, if present, have a negligible effect on the critical fluctuation of magnetization.

Above analyses on critical behaviors indicate the dominant long-range exchange interactions, while both the deviation from Curie-Weiss paramagnetic behavior above $T_c$ in a linear compound and previous DFT calculation suggest the important role of short-range exchange interaction in $Gd_6FeBi_2$ for high $T_c$. Actually the $T_c$ of $Gd_6FeBi_2$ is much higher than that of its analogue $Gd_6CoTe_2$ (350 K against 220 K) [31], which is not expected by RKKY interactions.

It is also known that the ferromagnetism/ferrimagnetism-paramagnetism transition is actually the destabilization of magnetic orders by thermal coupling. The scaling equation of state actually involves the role of thermal energy. Since $h_0$ is the effective exchange interaction field, the product of $h_0$ and an average effective elementary moment ($\mu_{eff}$) involved in the FM-PM transition, namely the effective exchange energy $\mu_{eff} h_0$, is expected to equal the thermal energy at $T=T_C$. In mean-field model, $\mu_{eff} h_o/k_B T_C$ equal 1.73. Using the value, $\mu_{eff}$ is estimated as $4.55\mu_B$ (~ $6.8\mu_B$ per Gd atom) for the $Gd_6FeBi_2$ compound, smaller than the theoretical value (7 $\mu_B$ per Gd atom) and suggestive of the Gd-Fe ferrimagnetic exchange interaction.

To address the inconsistence between critical exponents and short-range exchange interaction, it is necessary to analyse the structural origin of the Gd-Fe exchange interaction first. Like other $RE_6FeBi_2$, also present multiple magnetic transitions, and no evident structural change is observed at low temperature (see Fig 1). Neutron diffraction experiments show that these transitions are related to nonlinear magnetic structures. Unlike other *RE*, Gd is a linear alloy and does not show orbital-spin couplings for conical or helical magnetic structures, and thus the multiple magnetic transitions in are very likely to be related with the crystal-field effects. The typical energy between the ground state and the first excited crystal field state in *RE*s varies from 1 to 10 meV (10 K to 100 K) with the overall spread of all levels up to 20 meV [32], a value compared to the exchange energy for Gd (25 meV).

Because the magnetic structures of $Ho_6FeBi_2$ determined by neutron diffraction show no magnetic contribution from Fe [6], different structural features for the Gd-Fe ferrimagnetic interaction are expected in $Gd_6FeBi_2$. The electronic configure of Fe is $[Ar]3d^64s^2$, which allows a minimum-energy state with paired d electrons corresponding to a nonmagnetic state. However, the spatial extent of $3d$ electron wave function is considerably large, and thus $3d$ electron wave functions of neighbouring atoms show a strong overlap, leading to $3d$ electron bands, where the relatively strong effective Coulomb repulsion between 3d electrons can favor the situations in which the number of spin up and spin down electrons is no longer equal and leads to the formation of a magnetic moment. In the $RE_6TMX_2$ family, *TM* atoms are surrounded only by *RE* atoms (Fig 5a). The absence of *TM-TM* contacts and the low electronic density of states of *s* electron bands in *RE*s account for the nonmagnetic state of *TM* in these compounds except Gd analogues. In Gd analogues, the $5d$ electron band of Gd may have an overlap with $3d$ electron bands in *TM*, depending on the

Gd-*TM* bond strength. In $Gd_6CoTe_2$, for instance, the Gd-Co interaction seems similar in strength to other *RE*-Co as characterized by the nearly linear relationship between a/c and atomic radii of *RE* elements (Fig. 5b), because lattice constant a is related to the Fe site and a/c is an indicator of the *RE*-Fe bond strength. In contrast, Gd-Fe interaction is much stronger than other *RE*-Fe interactions in $RE_6FeBi_2$, which is characterized by the a/c of $Gd_6FeBi_2$ much smaller than the extrapolated value from the linear relationship between a/c and atomic radii of *RE* elements in other $RE_6FeBi_2$. Thus a stronger overlap of electron wave functions between Gd 5*d* electron bands and Fe 3*d* electron bands is expected, accounting for the magnetic moment from Fe in $Gd_6FeBi_2$. Such strong Gd-Fe interaction also benefits the spin polarization of Gd 5*d* electrons for an extra magnetic moment, in consistent with DFT calculations. In $Gd_6CoTe_2$, however, the overlap should be very weak or negligible due to the weak Gd-Co interaction, and a net magnetic moment from Co may not be expected.

The interactions between Gd and Fe in $Gd_6FeBi_2$ can be inspected more directly from its electronic structure, as seen in the XPS spectra in Fig. 6. Compared with sharp peaks in their electronic density of states (DOS) around the fermi level $E_F$ in Gd and Fe metals, $Gd_6FeBi_2$ has a more flat DOS up to several eV below $E_F$ (Fig 6a), suggestive of Gd-Fe hybridization and in agreement with previous DFT calculations. The low and flat DOS at $E_F$ supports the high stability of $Gd_6FeBi_2$, because the orbital splitting due to the hybridization allows more low-energy bands. However, the DOS of $Gd_6FeBi_2$ at $E_F$ is still much larger than that of $Gd_4Bi_3$ [14], a compound with a high $T_c$ (340 K) due to an extra Bi magnetic moment induced by the strong Gd-Bi hybridization [17]. Thus the magnetic contributions from Gd 5*d* and Fe 3*d* are expected in $Gd_6FeBi_2$, despite the reduced magnetic moment from *d* electrons due to some unpaired *d* electrons in these elements moved into lower energy orbitals to pair other electrons after the hybridization.

Compared with the strong overlap of *d* electron wave functions between neighbouring atoms in Gd and Fe metals, the overlap is effectively reduced in $Gd_6FeBi_2$ due to the hybridization and low DOS around $E_F$. Therefore the exchange interaction between Gd 5*d* and Fe 3*d* $Gd_6FeBi_2$ in should be weaker than those of Gd 5*d* –Gd 5*d* in Gd and *TM* 3*d* – *TM* 3*d* in Fe, which accounting for critical exponents more close to the values predicted by the long range interactions. In other words, the critical exponents corresponding to the Ising or Heisenbergy model in *Gd-TM* alloys should result from the *TM-TM* exchange interaction, because the Gd-*TM* exchange interaction is not strong enough. Thus the critical behavior of *Gd-TM* alloys can be tuned from the Ising/Heisenbergy model to the mean-field model by hybridizing *TM* with a third element (e.g. metalloid elements). This is exactly the case of $Gd_4Co_3$ alloys doped by Si: from $\beta$ = 0.389 and $\gamma$ = 1.229 in $Gd_4Co_3$ to $\beta$ = 0.465 and $\gamma$ = 1.134 in $(Gd_4Co_3)_{0.9}Si_{0.1}$ [33].

The ordering temperature in these compounds should be related to not only the magnetic moment, but also the vibronic coupling effects. Due to the strong Gd-Fe hybridization, the influence of such coupling on *d* electrons in $Gd_6FeBi_2$ should be much smaller than those of the Gd metal, and thus the short-range exchange interaction in $Gd_6FeBi_2$ diminishes more slowly with increased temperatures, accounting for the elevated $T_c$ and the deviation from the Curie-Weiss paramagnetic behavior above $T_c$. Similar phenomena are also observed in Si-doped $Gd_4Co_3$ with an increased $T_c$ (213 K against 208 K in $Gd_4Co_3$) [33].

Compared with Gd, the 4*f* band of $Gd_6FeBi_2$ has a lower energy state (more below the $E_F$) and show a clear splitting feature as a results of the hybridization and crystal-field effect (Fig 6b), suggestive of the higher structural stability. Although it does not produce extra moments, the split peak shifting towards low energy and spin-spin interactions should make it stronger against the vibronic coupling and thus also contribute to an elevated $T_c$. Such splitting of majority bands seems to be a common feature of electronic structures in the Gd-NME alloys with a high $T_c$, such as $Gd_5Si_4$, $Gd_4Bi_3$ and the present compound [11, 17, 34]. It is also suggested by the observation that the temperature shift of the vibronic peak was proportional to the splitting of ground level in LaCaMnO perovskite manganites [35]. However, the splitting also implies a wider temperature region of the magnetic transition or a small slope of moment against temperature, leading to small magnetic entropy change as confirmed by experiments on these alloys [34, 36, 37].

### 4.3 Crystal-field effects

According to above discussion, the crystal-field effect related to the Gd-Fe hybridization also plays an important role for the high $T_c$, and thus a brief analysis is needed here. The local atomic environments (AEs) around two kinds of Gd atoms in $Gd_6FeBi_2$ can be representatively depicted as Gd-centered polyhedra, as shown in Fig 7a and 7b. It is clear that Gd1 at $3g$ site has a octahedral symmetry marked as $O_h$, while Gd2 at $3f$ site shows a $C_{4v}$ symmetry in the form of square pyramid. For the $O_h$ symmetry, the sevenfold orbital degenerate F item of the free Gd atom with one $f$ electron in each orbital is split by the octahedral crystal field into three terms as depicted in Fig 7c: one is nondegenerate ($A_{2u}$) and two are threefold degenerate ($T_{1u}$ and $T_{2u}$ respectively) [38,39]. The splitting suggests DOS spectra of two main peaks of high energy levels plus a narrow peak of low energy levels. For the $C_{4v}$ symmetry, the F item is split into five terms (three nondegenerate items $A_1$, $B_1$, $B_2$ and two twofold degenerate items $E_1$ and $E_2$) by the ligand field of square prism, as shown in Fig 7d [38]. Generally, the energy gap of the splitting F →$A_1$+$B_1$+$B_2$+ $E_1$+ $E_2$ is small, thus one or two broadening $4f$ peak instead of five peaks was expected. In the splitting F →$A_{2u}$ + $T_{1u}$ + $T_{2u}$, the energy gaps between T items and F is small compared with that between A and F, and seems similar to that in the splitting of $C_{4v}$ symmetry. Thus a broadening 4$f$ peak including T, B, E and $A_1$ items is expected, while the $A_{2u}$ can be also enclosed or form a split peak. If $A_{2u}$ formed a split peak, the ratio of the small peak to the main peak should be around 1:13 in intensity. The inference was well supported by the observation of a shoulder in the broad Gd 4$f$ XPS peak (Fig 6b). It should be noted that the energy gap of band splitting is highly dependent on the strength of crystal fields and shortly expressed as [38]

$$\Delta = 5/3 eqF(R) \quad (17)$$

where $e$ is electronic charge, $q$ atomic effective charge, $R$ is interatomic distance and $F$ is a function of $R$. Thus the isostructural compounds can show different electronic structure, e.g. larger splitting energy which can even induce the filling of minor bands, or no observed splitting peak which corresponds for a relatively lower curie temperature. In the viewpoint here, the low $T_c$ in the isostructural $Gd_6CoTe_2$ can be well understood[31]. In turn, the energy gap of the splitting also is an indicator characterizing the stability of the crystal structure.

## 5. Conclusions

In conclusion, the critical behaviors of $Gd_6FeBi_2$ at $T_c$ are investigated comprehensively by isothermal magnetization, and estimated critical exponents are more close to the values predicted by the mean-field model. The analyses based on its crystal and electronic structure show that the Gd-Fe exchange interaction is much weaker than Fe-Fe exchange interaction and thus its influence on the critical behavior is limited, which is likely to be a typical feature for *RE-TM* alloys without *TM-TM* exchange interactions. Due to the strong Gd-Fe hybridization, the influence of vibronic couplings on the Gd-Fe short-range exchange interaction is diminished, accounting for the high $T_c$. On the other hand, the crystal-field effect results in the broadening or splitting of $f$-electron bands, and provides another important insight for the elevated $T_c$ in Gd-NME compounds and remarkably different magnetic behaviors among isostructural compounds.

**Acknowledgements**

The work was supported by National Key R&D Program of China (Grant No. 2016YFE0204200) and also partially supported by a grant from the Research Grants Council of Hong Kong SAR [(RGC Project No.) CityU 11253716]. We thank Dr. Z.G. Zheng in South China University of Technology for useful discussions on critical behaviors.

Table. 1 Comparison of critical exponents of Gd6FeBi2 and different theoretical models.

| composition | Technique | $\beta$ | $\gamma$ | Ref. |
|---|---|---|---|---|
| $Gd_6FeBi_2$ | Modified Arrott Plot | 0.441(8) | 1.098(12) | present |
|  | Kouvel-Fisher method | 0.446(6) | 1.092(10) | present |
| 3D Ising |  | 0.325 | 1.241 | [26] |
| 3D Heisenberg |  | 0.365 | 1.386 | [26] |
| Mean Field |  | 0.5 | 1 | [23] |

**Figure Captions**

**Fig. 1.** Experimenal XRD patterns (black) of $Gd_6FeBi_2$ and the simulated pattern (red) from the crystal structure determined at 150 K.

**Fig. 2.** Arrott plots (a) and modified arrott plots using different exponents: (b) 3D Ising model, (c) 3D Heisenberg model and (d) $β=0.441(8)$, $γ=1.098(12)$.

**Fig. 3.** (a) Temperaturea variation in spontanous magnetization $M_S(T)$ (left axis) and inverse initial susceptibility $χ_0^{-1}(T)$ (right axis); (b) Kouvel-Fisher plot of spontanous magnetization $M_S(T)$ (left axis) and inverse initial susceptibility $χ_0^{-1}(T)$ (right axis). Black solid lines in (a) are guides for eyes, and red straight lines in (b) are due to linear fitting of data.

**Fig. 4.** (a) Scaled magnetization of the $Gd_6FeBi_2$ compound below and above $T_C$, using critical exponents given by the Kouvel-Fisher plot. (b) The scaled magnetization and field are plotted in the form of $m^2$ vs. h/m for the $Gd_6FeBi_2$ compound. The inset in (a) shows the same plot on a log-log scale.

**Fig. 5.** (a) The representative crystal structure of $Gd_6FeBi_2$ viewed along c axis, and (b) plots of a/c against atomic radii of *RE* elements in $RE_6FeBi_2$ (black) and $RE_6CoTe_2$ (blue). The dashed circle in (b) emphasizes the contraction in the basal plane of $Gd_6FeBi_2$.

**Fig. 6.** XPS spectra of $Gd_6FeBi_2$ and the constitute elements in the valence band (a) and Gd 4*f* region (b). The intensity in XPS spectra of constitute elements is multiplied by their fraction in the formula.

**Fig. 7.** Atomic environments of Gd atoms in $Gd_6FeBi_2$: (a) Gd1 atom and (b) Gd2 atom, and schematic ilustration on corresponding crystal-field spliting of energy bands for Gd1 (c) and Gd2 (d).

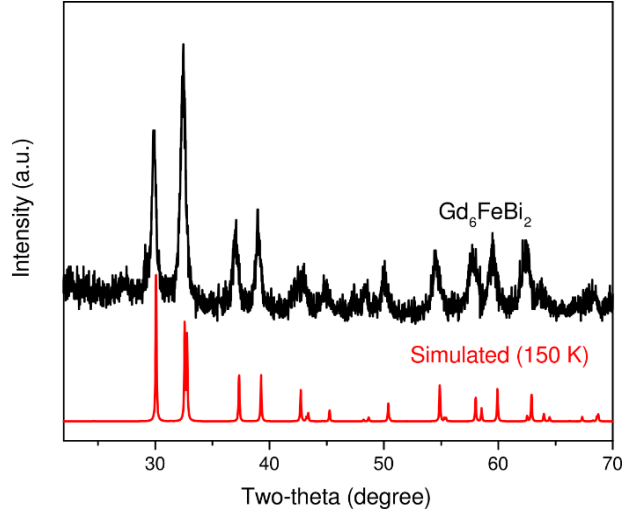

**Fig. 1.**

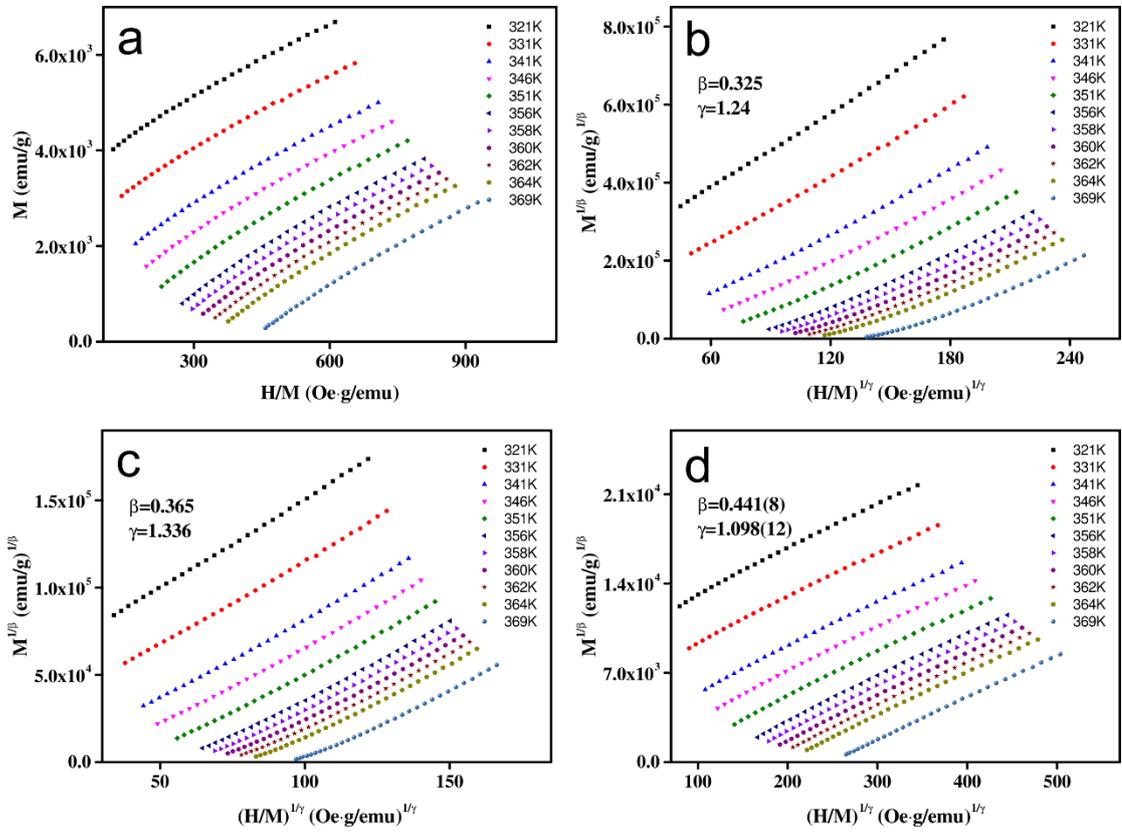

**Fig. 2.**

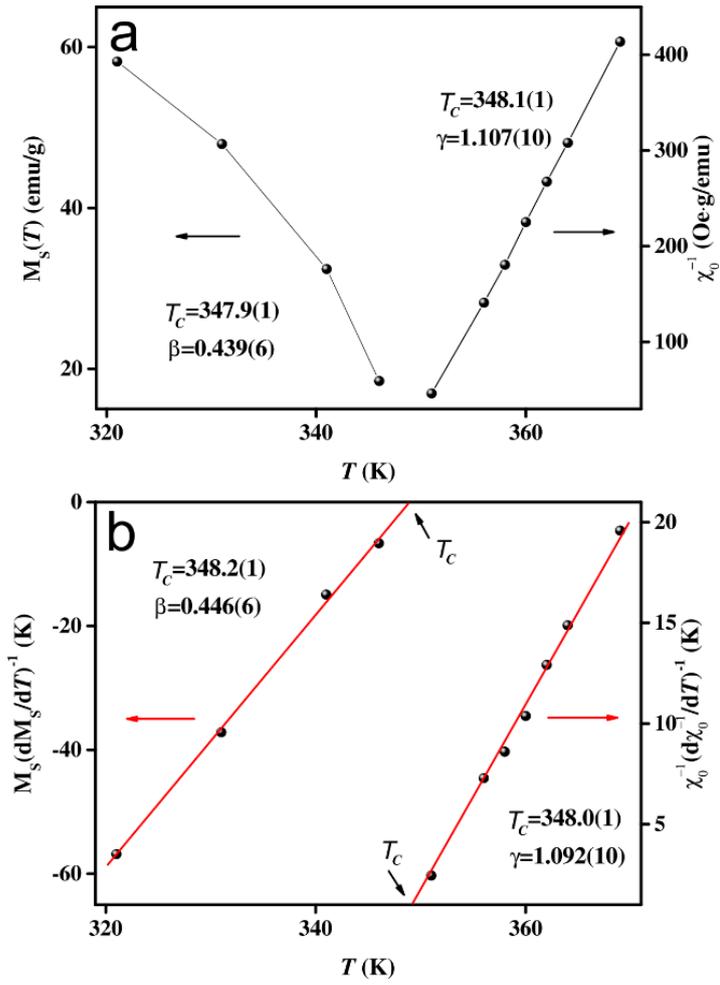

**Fig. 3.**

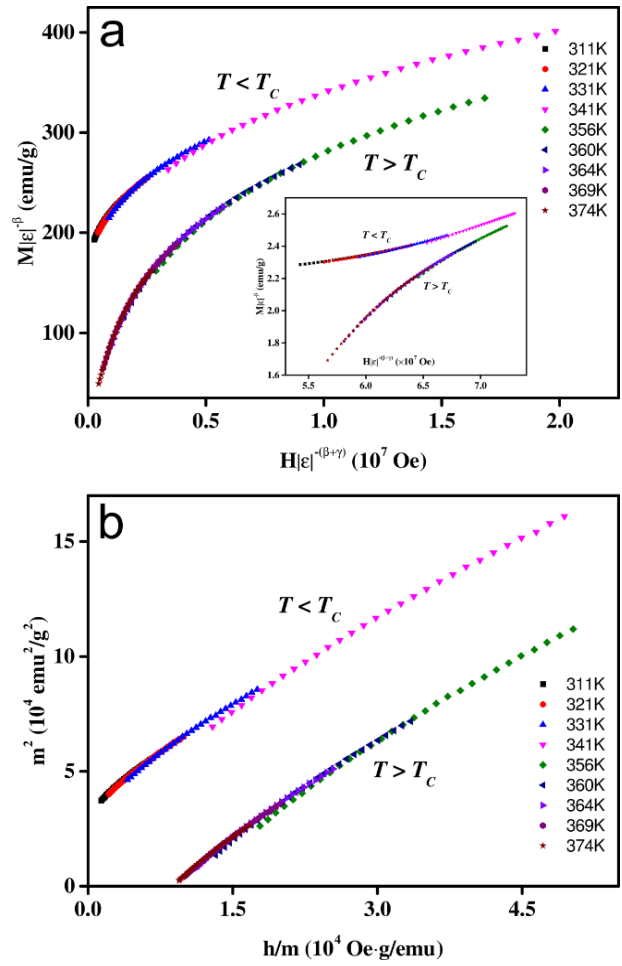

**Fig. 4.**

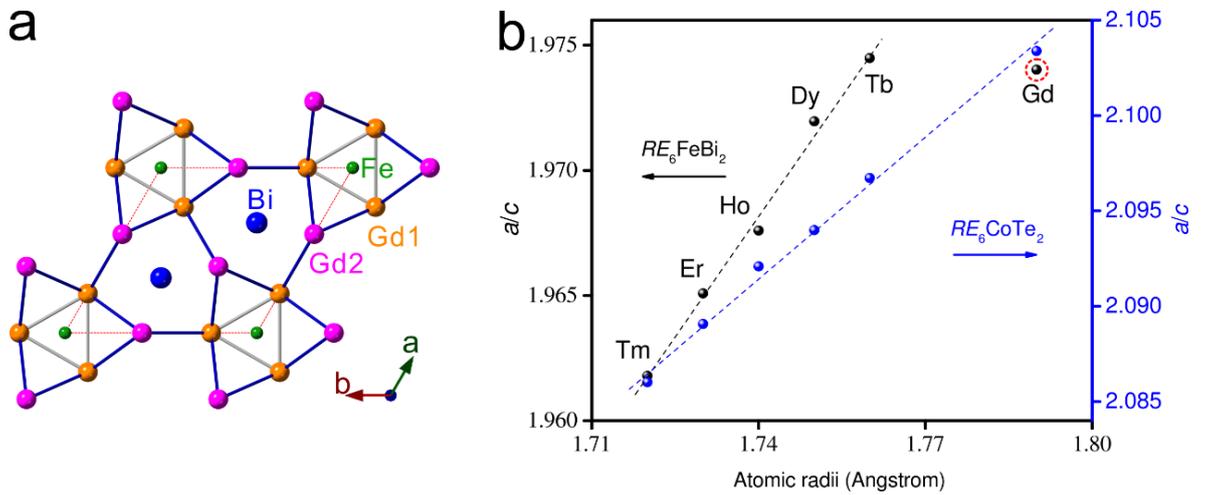

**Fig. 5.**

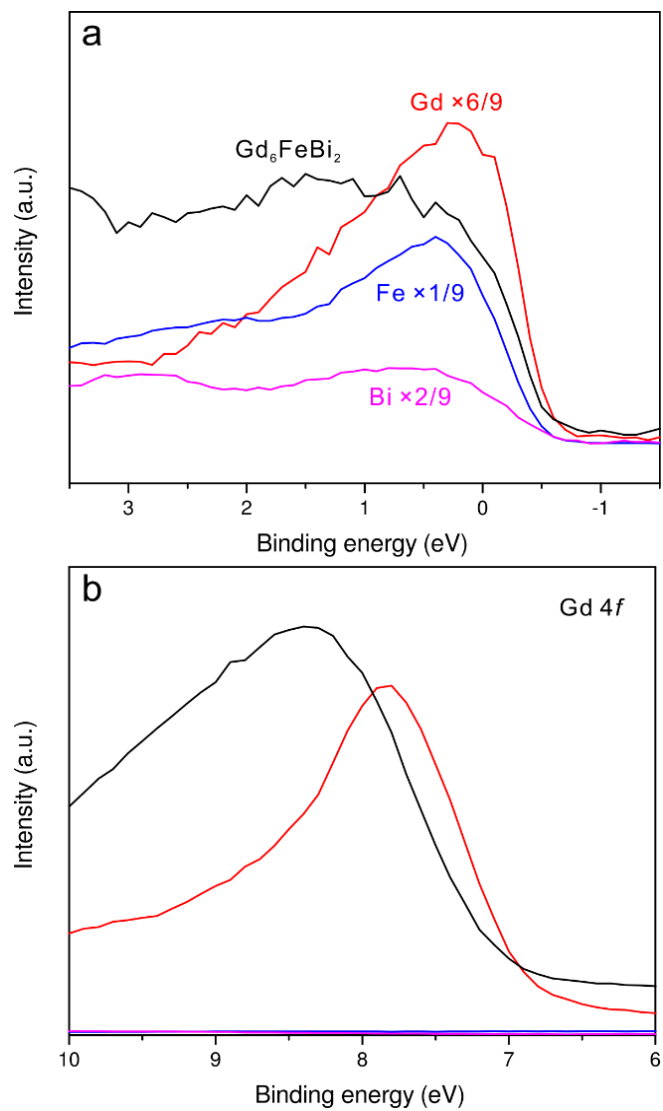

**Fig. 6.**

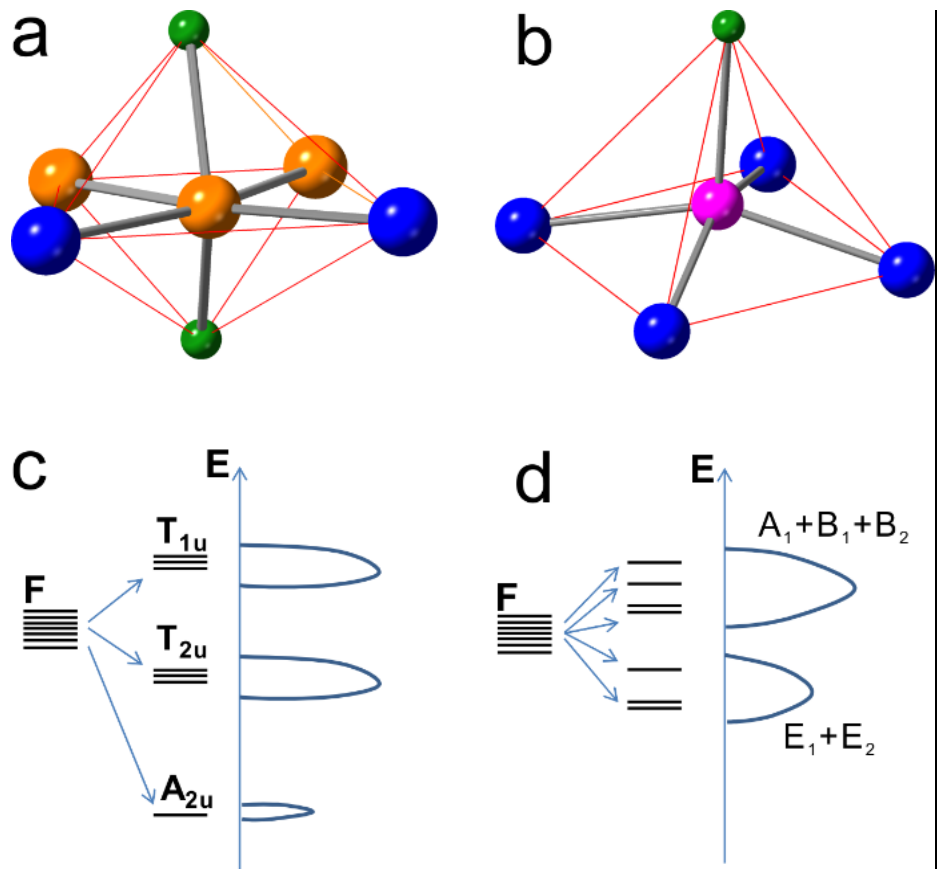

**Fig. 7.**